%% file: aa_consistency.tex
\documentclass[letterpaper, 11pt]{article}
\usepackage{appendix}
\usepackage{multirow}
 
\newcommand{\comment}[1]{}
 
\usepackage{graphicx}

\usepackage{subfigure}
\usepackage{amsmath}
\usepackage{amssymb}
\usepackage{mdwlist}

\usepackage{hyperref}

\usepackage{xspace}
\usepackage{setspace}

\newcommand{\sv}{{\mathcal V}}
\newcommand{\se}{{\mathcal E}}

\newcommand{\sP}{{\mathcal P}}
\newcommand{\precede}[2]{\xrightarrow[{\bf #2}]{#1}}

\newtheorem{definition}{Definition}

\def\noflash#1{\setbox0=\hbox{#1}\hbox to 1\wd0{\hfill}}

\begin{document}
\title{Application-Aware Consistency: An Application to Social Network \footnote{\normalsize This research is supported in part by National Science Foundation awards 1409416. Any opinions, findings, and conclusions or recommendations expressed here are those of the authors and do not necessarily reflect the views of the funding agencies or the U.S. government.}}

\author{Lewis Tseng$^{1}$, Alec Benzer$^{1}$ and Nitin H. Vaidya$^{2}$\\~\\
 \normalsize $^1$ Department of Computer Science,\\
 \normalsize $^2$ Department of Electrical and Computer Engineering, 
\\ \normalsize University of Illinois at Urbana-Champaign\\~\\ \normalsize Email: \{ltseng3, benzer2,
nhv\}@illinois.edu \\~\\} 

\date{Feb. 15, 2015\footnote{\normalsize Modified in May, 2015 to add more citations.}\footnote{\normalsize Modified in Oct., 2015 to improve presentation.}}
\maketitle

~

\begin{abstract}
\normalsize
This work weakens well-known consistency models using graphs that capture applications' characteristics. The weakened models not only respect application semantic, but also  yield a performance benefit. We introduce a notion of \textit{dependency graphs}, which specify \textit{relations} between events that are important with respect to application semantic, and then weaken traditional consistency models (e.g., causal consistency) using these graphs. Particularly,
we consider two types of graphs: \textit{intra-process} and \textit{inter-process} dependency graphs, where intra-process dependency graphs specify how events in a single process are related, and inter-process dependency graphs specify how events across multiple processes are related. Then, based on these two types of graphs, we define new consistency model, namely {\em application-aware} consistency.
We also discuss why such application-aware consistency can be useful in social network applications. 


\end{abstract}

\thispagestyle{empty}
\newpage
\setcounter{page}{1}

\input{intro}

\input{model}

\input{DGB}

\input{FGB}


\bibliography{paperlist,newbft,BFTnitin}

\appendix
\input{appendix}

\end{document}

%% file: intro.tex
\section{Introduction}
\label{s:intro}

Geo-replicated storage is commonly used to store data nowadays \cite{Azure,Cassandra,Spanner}. Replication brings with it the problem of maintaining \textit{consistency} across the replicas. Strong notions of consistency can be \textit{expensive} to achieve, often resulting in a non-trivial increase in latency in performing the operations. The difficulties of implementing replicated storage were captured in the CAP theorem \cite{CAP_Brewer,CAP_proof}, which essentially says that strong consistency, availability, and partition tolerance cannot {\em all} be achieved simultaneously in a replicated system. As observed in \cite{abadi-pacelc,codahale}, partition tolerance is necessary in many practical systems, and thus, cloud service providers have often chosen to support replication under {\em weaker} notions of consistency (e.g., \cite{Cassandra,Dynamo,Causal+,Red_Blue,terry13sosp,lazy_replication,lazy_replication_PODC}). 

Theses {\em weaker} consistency models -- although motivated by applications' need for
partition-tolerance -- are often agnostic of other application
characteristics or application semantics. For instance, the {\em causal consistency} model \cite{causal_memory} ensures that causal order \cite{lamport_causal} is enforced on 
events observed by any user, regardless of the \textit{relations} between events.
We will use the terms events and operations interchangeably. In many applications, particularly
social networking, the causal order may be too strong, and thus cause unnecessary latency \cite{Bailis_SOCC12,Bailis_blog}. To reduce such unnecessary latency, it is possible to weaken causal consistency without compromising on user-perceived ``quality'' of the information,
by taking into account the applications' semantics. 
We propose using dependency graphs that describe important application-specific ordering constraints, and relax the ordering constraints that are not important. This weakening of ordering constraints (when compared to many existing consistency models) can potentially result in better performance.  Ladin et al. also proposed a model allowing users to define important dependencies by using lazy replication \cite{lazy_replication,lazy_replication_PODC}. The relationship of our approach to Ladin et al. is discussed later in Section 2.
For brevity, we will only focus on weakening causal consistency \cite{causal_memory}, and addressing why such weakened causal consistency is adequate for social network applications. Section \ref{ss:intra} briefly discusses how to relax other consistency models using our approaches. We believe that these relaxed consistency models can be useful in some other applications. 


\paragraph{Motivation:}

Consider a simplified version of Facebook-like application: each user has a \textit{wall} where users can add new posts or comment on old posts. Later, Section \ref{ss:DGB_application} will address more complex operations, such as adding or removing friends. The application is implemented on top of a geo-replicated storage system, which supports two operations -- read and write, and stores data on multiple geographically distributed
replicas. The users access each other's posts from the closest
available replica. Similarly, a user's posts (or writes) are first propagated to one of
the replicas, which in turn, will propagate them to other replicas. 
The \textit{consistency model} being supported determines how the posts are propagated to the replicas, and when the posts become {\em visible} to users.

Facebook-like applications are usually implemented on top of systems ensuring {\em eventual consistency} \cite{Cassandra,Cassandra_paper} or {\em causal consistency} \cite{Causal+,bolt-on} due to smaller latency and the ability to work in partitioned network \cite{UTexas_tech_CAP}. However, these two models have their own drawbacks. On one hand, as discussed in \cite{Causal+,Causal+_ACMQ}, eventual consistency does not respect application semantic. Consequently, users may observe undesirable outcomes. On the other hand, causal consistency model may result into unnecessary latency for some events due to its restrictive ordering constraints \cite{Bailis_blog,Bailis_SOCC12}. Appendix \ref{a:drawback} elaborates on the limitations of eventual and causal consistency.


In short, we need a new consistency model for social network applications. Our solution is {\em application-aware} causal consistency, which weakens the causal ordering constraint based on \textit{dependency graphs} that specify applications' characteristics. First, {\em application-aware} causal consistency is based on causal consistency, and hence, unlike eventual consistency, it will respect the necessary application semantic. Second, by weakening the causal ordering, we \textit{reduce} the number of dependent operations for each operation without compromising on user-perceived ``quality'' of the information. 
This weakening technique can result in a \textit{reduced latency} in
completing operations. In particular, the delay before an operation can be made visible
to a client can be smaller, because dependency of a {\em smaller subset of events}
must be enforced (i.e., the operation needs to wait on a smaller number of prior operations
before it can be propagated). Waiting for fewer operations (or messages) implies that the expected delay
is typically smaller, due to delay variability. 

\begin{figure}[hbtp!]
\centering
\begin{minipage}[b]{0.48\linewidth}
\includegraphics[scale=0.55]{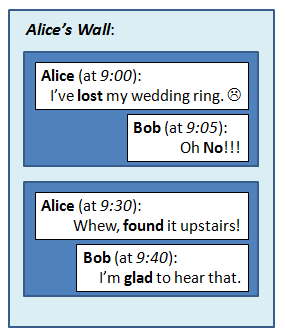} 
\caption{\it In the discussion, we use ``{\bf lost}'' to represent Alice's post at 9:00, ``{\bf no}'' for Bob's comment at 9:05, ``{\bf found}'' for Alice's post at 9:30, and ``{\bf glad}'' for Bob's comment at 9:40, respectively.
}
\label{f:alice}
\end{minipage}
\quad
\begin{minipage}[b]{0.48\linewidth}
\includegraphics[scale=0.55]{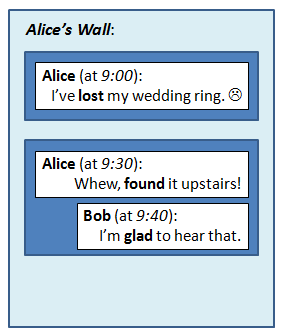}
\caption{\it This example illustrate that missing comments do not affect users' understanding. \newline\newline\newline
}
\label{f:alice2}
\end{minipage}
\end{figure}

In the discussion below, we will use an example in Figures \ref{f:alice} and \ref{f:alice2} to illustrate our relaxed causal consistency models. The example is adapted from an example presented in \cite{Causal+_ACMQ}. Consider three users, Alice, Bob and Calvin, who interact with each other via the social network application -- assume that all three of them can indeed access each other's wall because they are all ``friends'' in the social network. The posts (by Alice) and comments (by Bob) on Alice's wall and the corresponding timestamps are shown in Figure \ref{f:alice}. In our discussion, we will use the term {\em post} to refer to the text appearing first for a certain \textit{topic}, and the term {\em comment} for the text ensuing some {\em posts}. For example, Alice's update on 9:00 is a post, and Bob's update on 9:05 is an ensuing comment. For the example in Figure \ref{f:alice}, we will use ``{\bf lost}'' to represent Alice's post at 9:00, ``{\bf no}'' for Bob's comment at 9:05, ``{\bf found}'' for Alice's post at 9:30, and ``{\bf glad}'' for Bob's comment at 9:40, respectively.

\paragraph{Weakening Techniques:}
We propose two orthogonal methods to define \textit{application-aware causal consistency}. The core idea is to reduce the number of dependency operations (defined by the causal consistency) for each operation without violating application semantic. As addressed above, this weakening will likely result in a reduced latency.
\begin{itemize}
\item \textit{The first approach considers how the events in a single process are related:}

This is motivated by the following observation: \textit{the ordering observed by a single user may not necessarily reflect the application semantic; thus, the sequential ordering for a user's operation (program order) does not need to be respected at all time.}
Consider the example in Figure \ref{f:alice}. From Alice's perspective, ``{\bf found}'' does not causally depend on ``{\bf no}''. In other words, even though Alice reads ``{\bf no}'' before posting ``{\bf found}'', ``{\bf no}'' does not {\em cause} Alice to post ``{\bf found}''. Consequently, it is fine if Calvin observes ``{\bf found}'' before observing ``{\bf no}''. That is, even if Calvin observes Alice's wall as shown in Figure \ref{f:alice2}, Calvin can still follow the whole story. 
Such a weakening may be desirable if the propagation delay of ``{\bf no}'' is too large. In this case, even when Calvin's replica already had received ``{\bf found}'', the replica cannot make ``{\bf found}'' visible to Calvin (for satisfying causal order \cite{lamport_causal}). As a result, Calvin has to wait for the long delay of ``{\bf no}'', which is an unnecessary delay. 

In short, we want a consistency model that achieves two following goals: (i) allowing Calvin to observe ``{\bf found}'' before ``{\bf no}'' to shorten latency, and (ii) requiring Calvin to observe ``{\bf found}'' before ``{\bf glad}'' to enforce application semantics. It is obvious that these goals cannot be simultaneously satisfied by enforcing either causal or eventual consistency. 
As a remedy, we propose {\em application-specific} consistency based on the notion of \textit{intra-process dependency graph}, which is a directed acyclic graph describing how each operation is \textit{related} to other operations at a single process. For instance, in Calvin's scenario, the dependency graph will specify that there is no relation between ``{\bf found}'' and ``{\bf no}''. Thus, Calvin's replica can make ``{\bf found}'' visible to Calvin even if Calvin has not yet seen ``{\bf no}''. Effectively, the intra-process dependency graph allows us to relax {\em program order} without violating important application semantic. Section \ref{s:intra} discusses this method in details. In particular, it shows that
the new causal consistency model based on the intra-process dependency graph achieves the two aforementioned goals.  Note that Ladin et al. proposed a similar model in \cite{lazy_replication,lazy_replication_PODC}. We discuss the related work in more detail in Section \ref{s:related}.

\item \textit{The second approach considers how the events across multiple processes are related:}

We use an \textit{inter-process dependency graph}, which explicitly describes how each read operation at a process is \textit{related} to other processes' write operations. One special form of inter-process dependency graph is the ``friends'' graph, in which two users are connected if they are each other's friends in the social network. Intuitively, each user C must observe operations performed by users within distance $d$ in the
``friends'' graph in the causal order {\em induced only by operations performed by users
within distance $d$}.
Operations performed by users farther than distance $d$ may be observed by C in an arbitrary order. We elaborate on the motivation using an example below.


\begin{figure}[htbp]
	\centering
	\begin{minipage}[b]{0.48\linewidth}
	\includegraphics[scale=0.48]{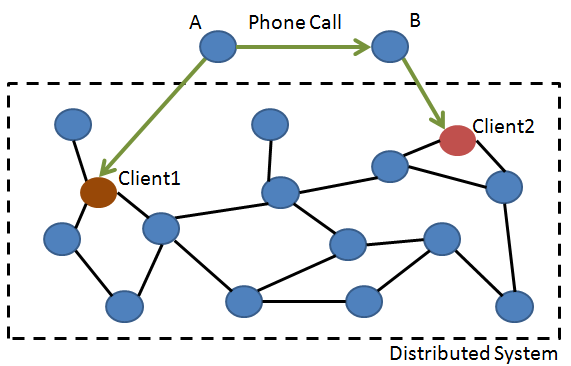}
	\caption{\it The view of universe in traditional causal consistency.
	}
	\label{fig:friend1}
	\end{minipage}
	\quad
	\begin{minipage}[b]{0.48\linewidth}
	\includegraphics[scale=0.48]{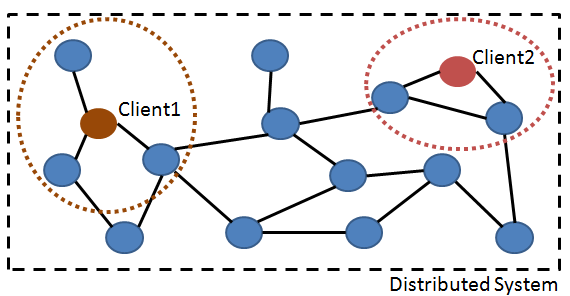}
	\caption{\it The view of universe in weakened causal consistency.
	}
	\label{fig:friend2}
	\end{minipage}
\end{figure}

Figure \ref{fig:friend1} shows a social network, with the graph in the figure depicting the
friends relation. In this system, it is possible that agents A and B that are external to the
social network may send messages (such as e-mails) to Client1 and Client2 (who are in the social network)
as shown in the figure. In fact, the message from agent B may follow a phone call from
agent A to agent B. Thus, there is a potential causal dependence between messages sent by A and B.
However, the consistency protocol used for the social network has no way to take into account
this causal relationship. The clients find it acceptable that causality in the universe ``external''
to the social network is not always reflected in the behavior they observe. The friends graph-based consistency model
generalizes on the notion of external  universe.
In particular, as shown in Figure \ref{fig:friend2}, the notion of external universe
is defined {\em with respect to each client}. For Client1 and Client2, everyone outside $d$-hop
distance from them in the friends graph is viewed as external to their universe,
respectively.
Thus, everyone outside the dotted circle around Client1 is external to Client1's universe
for the purpose of enforcing causal delivery (and similarly for Client2).


We believe that such weakening of consistency is likely to be acceptable
in social networking contexts, provided that it yields a performance benefit. For instance,
the friends graph-based causal consistency model with
$d=1$ in the example in Figure \ref{f:alice} will still result in Calvin observing ``{\bf found}''
before ``{\bf glad}'', the desired behavior. Section \ref{s:inter} discusses this method. Note that this method shares some similarity to Fisheye consistency \cite{fisheye} and other consistency models using distance as a metric \cite{vector-field_consistency,Dagstuhl}. One novelty of using inter-process dependency graph is to allow different users to follow different ordering constraints. We discuss related work in detail in Section \ref{s:related}. 

\end{itemize}
For brevity, we will only focus on relaxing causal consistency using \textit{dependency graphs}, and discussing why our relaxed model is suitable for Facebook-like applications. Section \ref{ss:intra} discusses how to weaken other well-known consistency models.

\section{Related Work}
\label{s:related}

There is a long history of research on various weak consistency models
for parallel computers and for distributed shared memory systems, with significant activity on this topic
in the 1990s \cite{Adve_tutorial,Gharachorloo_consistency,lazy_release92}.
Recently, the weak consistency models have received renewed attention in the context of replicated storage systems, which aim to reduce access latency via {\em geo-replication}. 
With the users located across the globe, it has become important to keep the data close to all the users who share it, in order to reduce access latency. This has motivated {\em geo-replication}, or replication of data across geographically distributed replicas.
Such {\em geo-replicated} storage systems include Windows Azure \cite{Azure}, Cassandra \cite{Cassandra}, Amazon's Dynamo \cite{Dynamo} and Google's Spanner \cite{Spanner} and Megastore \cite{megastore}. 
Large latencies can still be incurred if the system insists on supporting
{\em strong notions of consistency}, since that effectively requires a total ordering of requests (or operations)
executed by the geographically distributed replicas. 
This has forced the system designers to address
the trade-off anticipated by the CAP theorem, which was initially conjectured by Brewer \cite{CAP_Brewer}, and later formally proved by Gilbert and Lynch \cite{CAP_proof}. 
Designers of many geo-replicated systems determined that {\em availability} and {\em partition-tolerance} are sometimes more important than {\em strong consistency} \cite{abadi-pacelc,hat-vldb2014}, and chose to relax the consistency guarantees (e.g., see \cite{Dynamo}).
Many weak versions of consistency have been proposed, such as eventual \cite{Cassandra,Dynamo}, causal+ \cite{Causal+}, RedBlue \cite{Red_Blue}, update \cite{update} and multi-level consistency (Pileus system) \cite{terry13sosp}. This line of work focuses more on availability-consistency trade-off in the face of partitioned network, and does not elaborate on frameworks to weaken existing consistency models. We discuss papers that share similarity with our work below.

Closest to our idea of using intra-process dependency graph is \cite{lazy_replication,lazy_replication_PODC}, which proposed a highly-available system that allows users to define dependencies among operations. While our motivation and consistency models are similar, there are several differences. First,  Ladin et al. described four rules for their consistency models without specifying how to apply their model in practice. For example, they did not consider how users may define dependencies for a given application, or how to learn the user-defined dependency in real time. On the contrary, our paper shows that given simple social-network application semantic, there exists a compact way to construct the dependency graphs, and thus, the consistency model is well-defined. Second,  Ladin et al. did not propose a general implementation for the consistency model proposed in \cite{lazy_replication,lazy_replication_PODC}. Instead, they implemented a system satisfying causal consistency \cite{causal_memory}. As discussed in our paper, the causal systems may be too strong in some scenarios. We are presently developing algorithms to implement the weakened models proposed in the paper. Please refer to \cite{thomas} for our preliminary implementation on the relaxed consistency model for a Facebook-like application. Finally, we propose a general framework to weaken previously proposed consistency models by relaxing those ordering constraints that are not important as defined in the dependency graphs. 

The idea of relaxing \textit{program order} at a process is not new. Prior work has also proposed similar ideas, e.g., eventual \cite{Dynamo,Cassandra_paper}, non-sequential \cite{nonsequential}, update \cite{update}, and RedBlue \cite{Red_Blue} consistency. Among this work, eventual \cite{Dynamo,Cassandra_paper} and update \cite{update} consistency did not consider application characteristics, whereas  \cite{nonsequential,Red_Blue} took into account application semantic. \cite{Red_Blue} divided operations into two categories: strong and weak, and the consistency models require strong operations to follow some stronger ordering constraints (e.g., sequential consistency \cite{lamport_sequential}) while weak operations may follow some more relaxed ordering constraint (e.g., PRAM \cite{PRAM} or eventual consistency). Non-sequential consistency \cite{nonsequential} allows users to observe operations in an order that does not respect (real-time) execution order. However, each user still observes the same total order. Their motivation is to provide guarantees on execution that may appear to be non-sequential due to speculative executions. Moreover, their system model is different from ours \cite{nonsequential}. They consider multiple processes accessing a set of physical registers, where as we consider shared memory built on top of multiple replicas. Therefore, in their system model, operations on the same key (or same register) are \textit{guaranteed to be atomic}; while in our system, operations on the same key may \textit{not} be atomic.

Prior work also attempts to relax ordering constraint of events across multiple processes. There is work using {\em distance} as a metric to define consistency models. \cite{Dagstuhl,vector-field_consistency} proposed geographical-distance-based consistency models for games, in which users observe nearby events in a strong ordering and far-away events in a weaker ordering. However, \cite{Dagstuhl,vector-field_consistency} did not consider usage of graphs, and it is not clear whether their consistency models can be easily extended to the case when distance becomes a virtual notion like the one in friends graphs.
Closest to our second approach (inter-process dependency graphs) is the recent work of
Friedman, Raynal and Ta{\"{\i}}ani on {\em Fisheye consistency}, which also incorporates the idea of using graphs to relax traditional consistency models \cite{fisheye}. Intuitively, Friedman et al. use a proximity graph (which is an {\em undirected graph}) to describe important ordering constraints.
While our approach
and the work on {\em Fisheye} consistency share the property of using graphs, there are some
important limitations to this early work.
The key shortcoming is that
Friedman et al. only present one concrete use of graphs, namely to support a
version of the {\em sequential} consistency model
for operations performed by nodes near each other (neighboring nodes in the proximity graph), and
causal consistency for operations performed by other nodes (non-neighboring nodes in the proximity graph).
They do not identify how to extend their
approach to weaken other useful consistency models using proximity graphs.
Most importantly, we allow the inter-process dependency graphs to be \textit{directed} graphs, while the proximity graph in \cite{fisheye} is assumed to be undirected. Later in Section \ref{ss:disc_inter}, we show that using \textit{directed} graphs allow us to capture richer application semantic.
The other main difference is that Friedman et al. do not consider how the operations in a single process are related (we use \textit{intra-process dependency graphs} to capture the relation of events occur at a process).

Bailis et al. also identified the scalability issue of ensuring causality \cite{Bailis_blog,Bailis_SOCC12}. \cite{Bailis_SOCC12} proposed tracking application-specific causal dependency, which results in higher write throughput due to faster check of causality dependency and smaller metadata. In his {\em blog} \cite{Bailis_blog}, Bailis also discussed several other methods to make causality cheaper, including sacrificing availability for more efficiently tracking causality. However, Bailis et al. did not specify how to track the dependency, nor did them provide a systematic way to relax existing consistency models in  \cite{Bailis_blog,Bailis_SOCC12}. Particularly, they did not consider usage of intra- and inter-process dependency graphs, and did not provide a formal definition of the relaxed consistency model in \cite{Bailis_blog,Bailis_SOCC12}. We share their motivation of improving performance by weakening the consistency requirements. Beyond \cite{Bailis_blog,Bailis_SOCC12}, we propose using dependency graphs that describe application-specific ordering constraints, and formally define relaxed consistency models based on the dependency graphs.

%% file: model.tex
\section{System Model and Terminology}
\label{s:terminology}


\paragraph{System Model:}

We consider a \textit{system} consisting of $n$ processes, $\sP = \{p_1, p_2, \cdots, p_n\}$. The geo-replicated storage is modeled as a shared memory of a finite set of objects . We will use lower case italic letters to denote the objects. Each process interacts with the shared memory through a series of read and write operations over a {\em reliable} channel. 
The system is assumed to be asynchronous. 
We further make the following assumptions to simplify our discussion: 

\begin{itemize}
\item No value is written more than once to the same object. This assumption simplifies the definition of causal order -- this can be relaxed by generalizing the reads-from order (presented below).

\item Each process is {\em sequential}, i.e., no two operations are executed at the same time. We do note that our model can be generalized to the case of multi-thread process. We will address this extension in Section \ref{ss:disc_intra}.

\item Each operation $o$ consists of an invocation event $inv(o)$ and a corresponding response event $resp(o)$. Moreover, each operation can access one object \textit{atomically} at a time. Generalizing our model to support transactions is a future research direction.
\end{itemize}


\paragraph{Terminology:}

We introduce the standard terminology of consistency models in shared memory, e.g., \cite{causal_memory,welch_book,AA_nancy}. 
For each process $p_i \in \sP$, the local history $L_i$ of
process $i$ is a sequence of read and write operations. 
If operation $o_1$ precedes $o_2$ in $L_i$, i.e., response event of $o_1$ occurs before invocation event of $o_2$ in $L_i$, we denote it by $o_1 \precede{L_i}{} o_2$.
A history $H = \{L_1, \cdots, L_n\}$ is the collection of local histories. 
A serialization $S$ of the history $H$ is a \textit{linear} sequence of all operations in $H$ in
which each read operation on an object $x$ returns its \textit{most recent} preceding write operation on the object $x$ (or $\perp$ if there is no preceding write in $S$). 
The serialization $S$ respects an order $\rightarrow$, if for any operation $o_1$ and $o_2$ in $S$, $o_1 \rightarrow o_2$ implies $o_1$ precedes $o_2$ in $S$. 

Given a history, the reads-from order $\precede{}{read}$ is defined as follows: if $o_r$ is a read that returns the value written by $o_w$, then $o_w \precede{}{read} o_r$. Recall that, for simplicity, we assume that no value is written more than once to the same object. This can be easily relaxed by generalizing reads-from order. Then, we define the causal order \cite{lamport_causal}. Given a history, $o_1 \precede{}{CC} o_2$, where $\precede{}{CC}$ represents the causal order, if any of the following holds \cite{lamport_causal}:

\begin{itemize}
\item {\bf Program-order}: $o_1 \precede{L_i}{} o_2$ for some $p_i$ ($o_1$ precedes $o_2$ in $L_i$),

\item {\bf Reads-from}: $o_1 \precede{}{read} o_2$ ($o_2$ returns the value written by $o_1$), or

\item {\bf Transitivity}: there is some other operation $o'$ such that $o_1 \precede{}{CC} o' \precede{}{CC} o_2$.
\end{itemize}

Let $H|(p_i+W)$ denote the set of all operations in the local history $L_i$ and all write operations in the history $H$. 

\begin{definition}
\label{def:CC}
{\bf (Causal Consistency \cite{causal_memory})} 
A history is {\em causally consistent} if for each $p_i \in \sP$, there exists a serialization $S_i$ of $H|(p_i+W)$ that respects the causal order $\precede{}{CC}$.
\end{definition}

We will discuss different methods to relax causal consistency. In particular, Section \ref{s:intra} relaxes the program-order rule, and Section \ref{s:inter} relaxes the reads-from rule. 

%% file: DGB.tex
\section{Intra-Process Dependency}
\label{s:intra}

Our first approach addresses using intra-process dependency graphs to relax consistency models. More precisely, we will generalize the program-order rule discussed in Section \ref{s:terminology}. Most consistency models, e.g., sequential \cite{lamport_sequential}, PRAM \cite{PRAM}, and causal consistency \cite{causal_memory}, linearizability \cite{linearizability}, enforce the existence of serialization(s) that respect the \textit{program order} at each process. However, as discussed in Section \ref{s:intro}, such requirement may not be necessary in some applications, since the ordering constraints defined by the program order may not reflect the real relations among operations. We first introduce the notion of \textit{intra-process dependency graphs} that characterize important ordering constraints based on the applications' semantic. Then, we weaken causal consistency based on the intra-process dependency graphs, thus the name -- \textit{intra-causal consistency}. Section \ref{ss:intra} discusses how to use dependency graphs to relax other well-known consistency models.

\textit{Intra-process dependency graph} captures the ordering of events at each process that are important and should be respected. For now, assume that the dependency graphs are given. Later in the section, we will briefly discuss how to generate intra-process dependency graphs for a Facebook-like applications.

\begin{definition}
\label{def:intra}
{\bf (Intra-Process Dependency Graph)} Given a history $H$, a graph for a process $i$, $D_i(\sv_i, \se_i)$, is an intra-process dependency graph if it satisfies the following properties: (i) Each node in $\sv_i$ corresponds to an unique operation in the local history of process $i$, $L_i$, and (ii) $D_i$ is a directed acyclic graph.
\end{definition}
With a slight abuse of terminology, we will use the operation name to refer to the node in $D_i$, the intra-process dependency graph at process $i$. Whether we are referring to the operation or the corresponding node will be clear from the context.

In essence, intra-process dependency graph for a process $i$ induces an \textit{intra-program order} (denoted as $\precede{L_i}{intra}$) -- which should be respected in consistency models. For brevity, we ignore the dependency graphs in the notation.

\begin{definition}
\label{def:intra-order}
{\bf (Intra-Program Order)}
Consider two operations $o_1$ and $o_2$ in the local history at process $i$, $L_i$, if node $o_1$ has a directed edge to node $o_2$ in $D_i$, then $o_1 \precede{L_i}{intra} o_2$.
\end{definition}

If $D_i$ is a chain of operations that follow the linear order of $L_i$, then $\precede{L_i}{intra}$ is identical to the program order $\precede{L_i}{}$.

\subsection{Intra-Causal Consistency}

We first show how to use intra-process dependency graphs to transform the traditional causal consistency into a new model -- \textit{intra-causal consistency}. 

Given intra-dependency graphs, a history $H$ induces an \textit{intra-causal order}, denoted as $\precede{H}{intra-CC}$. For simplicity, we ignore the intra-dependency graphs in the notation. $o_1 \precede{}{intra-CC} o_2$ if any of the following holds: 

\begin{itemize}
\item {\bf Intra-program-order}: $o_1 \precede{L_i}{intra} o_2$ for some $p_i$ (node $o_1$ has a directed edge to node $o_2$ in $D_i$),

\item {\bf Reads-from}: $o_1 \precede{}{read} o_2$ ($o_2$ returns the value written by $o_1$), or

\item {\bf Transitivity}: there is some other operation $o'$ such that $o_1 \precede{}{intra-CC} o' \precede{}{intra-CC} o_2$.
\end{itemize}

Recall that $H|(p_i+W)$ denotes the set of all operations in the local history $L_i$ and all write operations in the history $H$. 

\begin{definition}
\label{def:intra-CC}
{\bf (Intra-Causal Consistency)}
A history $H$ is {\em intra-causally consistent} if for each $p_i \in \sP$, there exists a serialization $S_i$ of $H|(p_i+W)$ that respects $\precede{}{intra-CC}$.
\end{definition}

Note that the difference between causal consistency and intra-causal consistency is in the first rule above. Whereas causal consistency is based on program order, we use \textit{intr-program order} for intra-causal consistency, which relaxes program order using the intra-dependency graph.

Intra-causal consistency is similar to the model proposed by Ladin et al. \cite{lazy_replication,lazy_replication_PODC}. In Ladin's system, they allowed user to define important dependencies, and use lazy replication to implement a system that respect these dependencies \cite{lazy_replication,lazy_replication_PODC}. Intuitively, \textit{intra-program-order rule} (based on intra-dependency graph) and \textit{reads-from rule} jointly specify the user-defined dependencies in Ladin's consistency model, and we can build a system based on lazy replication \cite{lazy_replication,lazy_replication_PODC}. However, in \cite{lazy_replication,lazy_replication_PODC}, Ladin et al. did not address how to identify these user-defined dependencies. To the best of our knowledge, we are the first to propose using graphs to capture these dependencies, and to address how to apply it to social network application.

\subsubsection{Application to Social Network}
\label{ss:DGB_application}

We use the scenario in Figures \ref{f:alice} and \ref{f:alice2} to illustrate why intra-causal consistency model is useful in social network applications. In particular, we discuss how intra-causal consistency achieves the following two goals: (i) allowing \textit{some} user to observe ``{\bf found}'' before ``{\bf no}'', and (ii) requiring \textit{each} user to observe ``{\bf found}'' before ``{\bf glad}''. As discussed in Section \ref{s:intro}, the first goal may reduce the latency observed by users, while the second goal ensures that ordering of events follows application semantic. However, these two goals cannot be achieved simultaneous by either eventual or causal consistency as addressed in Appendix \ref{a:drawback}. In the discussion below, we model each user of the Facebook-like application as a process. Thus, we will often use the terms user and process interchangeably.

\comment{+++++++++-
\begin{figure}[hbtp!]
\centering
\begin{minipage}[b]{0.48\linewidth}
\includegraphics[scale=0.45]{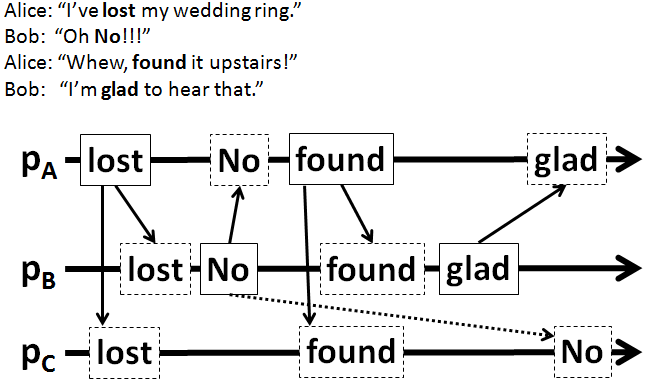} 
\caption{\it Real time interaction between Alice, Bob and Calvin. 
}
\end{minipage}
\quad
\begin{minipage}[b]{0.48\linewidth}
\includegraphics[scale=0.45]{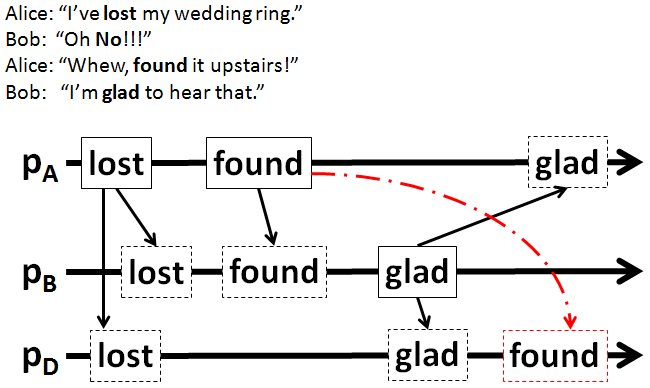}
\caption{\it Real time interaction between Alice, Bob and Darren. 
}
\end{minipage}
\end{figure}
++++++++++++++}

\begin{figure}[hbtp!]
\centering
\includegraphics[scale=0.75]{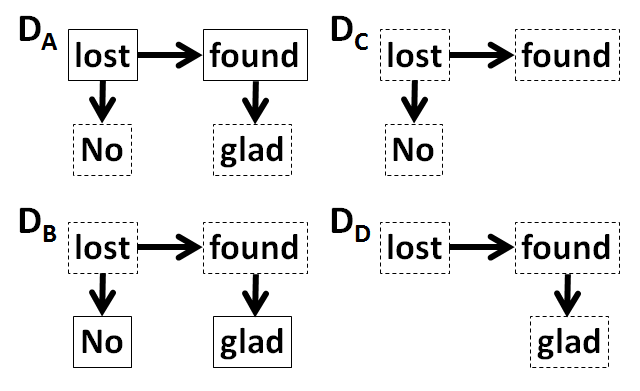}
\caption{\it Intra-process dependency graph for each user. Solid and dashed boxes represent write and read operations, respectively. For example, in the figure, the solid box of the text \textbf{lost} in $D_A$ represents the operation $w_A(x_{lost}){\bf lost}$ -- Alice writes message ``{\bf lost}'' (``I've {\bf lost} my wedding ring.'') to a distinct object $x_{lost}$.
}
\label{f:DGB_dependence}
\end{figure}

We assume that intra-process dependency graphs are given, which are shown in Figure \ref{f:DGB_dependence}. In the figure, the boxes with solid and dashed lines represent write and read operations, respectively. For simplicity, users and objects are not shown in Figure \ref{f:DGB_dependence}. Section \ref{ss:disc_intra} addresses how to construct the intra-process dependency graphs in an online fashion. We first discuss briefly how Facebook-like application is implemented. A write operation by an user will write a value to a {\em distinct} object, and propagate the value to all the replicas. A read operation can be related to reading snapshot of parts of the shared memory space, e.g., shared memory space storing data of Alice's wall in our example. Intuitively, each read returns the ``difference'' between snapshot returned by the previous read and the current snapshot (or empty snapshot if there is no previous read). We also borrow some standard notations \cite{causal_memory}: 

\begin{itemize}
\item $w_i(x_{msg}){\bf msg}$ denotes that user $i$ writes value ${\bf msg}$ to some distinct object $x_{msg}$. For example, $w_A(x_{found}){\bf found}$ means that Alice writes message ``{\bf found}'' (``Whew, {\bf found} it upstairs!'') to a distinct object $x_{found}$. 

\item $r_i(x_{Alice\_wall}){\bf msg}$ denotes that user $i$ reads the value ${\bf msg}$ -- which is the difference between snapshots of shared memory space that contains data of Alic's wall. In the figures below, we will use the returned values of the read to represent read operations. For instance, $r_A(x_{Alice\_wall}){\bf No}$ means that the read operations returns the value ``{\bf No}'' (``Oh No!!!'') at object $x_{No}$. This is a \textit{difference} between snapshots of Alice's wall, since after Alice wrote ``{\bf found}'', the snapshot at Alice's replica already contains the value ``{\bf found}''. Note that in the following figures, such a read operation is represented by a box with dashed line and text ``No''.
\end{itemize}


Now, we discuss why \textit{intra-causal consistency} achieves two aforementioned goals below.


\begin{itemize}
\item Allowing some user to observe ``{\bf found}'' before ``{\bf no}'':

Consider the real time interaction of the scenario shown in Figure \ref{f:abc}. In the figure, solid and dashed boxes represent write and read operations, respectively. Arrows denote the read-from relation.  For simplicity, some operations are not shown in these figures. 

\begin{figure}[hbtp!]
\centering
\includegraphics[scale=0.75]{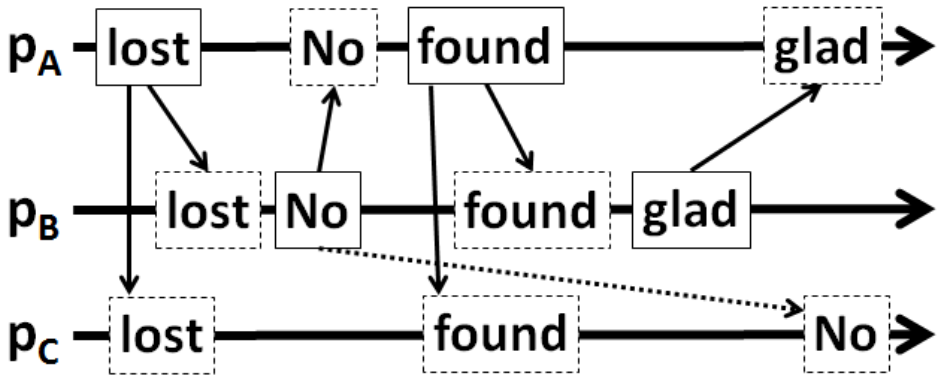} 
\caption{\it Real time interaction between Alice, Bob and Calvin. In the figure, solid and dashed boxes represent write and read operations, respectively. Arrows denote the read-from relation. For simplicity, some operations are not shown in these figures. 
}
\label{f:abc}
\end{figure}

Figure \ref{f:abc} shows a scenario that satisfies intra-causal consistency (the sequence of events shown in Figure \ref{f:abc} for each user is the desired serialization). Note that in the scenario, Calvin observes the posts in the order of ``{\bf lost}'', ``{\bf found}'', and ``{\bf No}''. This violates traditional causal consistency (Definition \ref{def:CC}), because if we look at the interaction of Alice and Bob, then $w_B(x_{No}){\bf No} \precede{}{CC} w_A{x_{found}}{\bf found}$; however, Calvin observes ``{\bf found}'' before observing ``{\bf No}''.
On the contrary, such a scenario is allowed under intra-causal consistency, since $w_A(x_{found}){\bf found}$ does not \textit{intra-causally depend} on $w_B(x_{No}){\bf No}$, i.e., we do not have $w_A(x_{found}){\bf found} \precede{}{intra-CC} w_B(x_{No}){\bf No}$ using three rules defined for intra-causal consistency (Definition \ref{def:intra-CC}).  

\item Requiring each user to observe ``{\bf found}'' before ``{\bf glad}'':

\begin{figure}[hbtp!]
\centering
\includegraphics[scale=0.75]{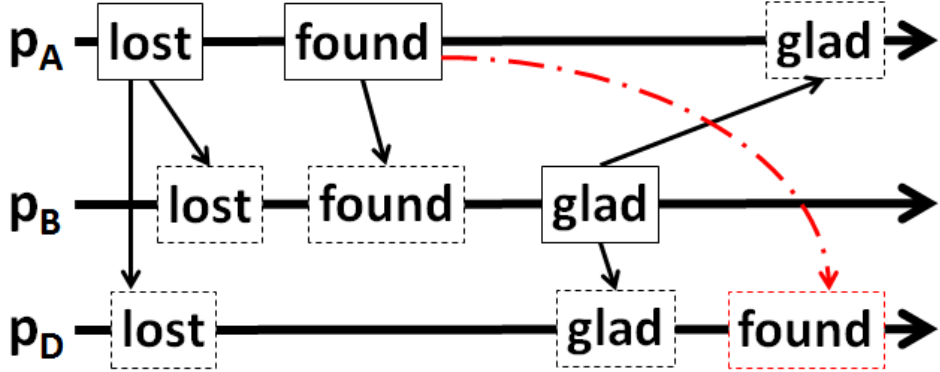} 
\caption{\it Real time interaction between Alice, Bob and Darren. In the figure, solid and dashed boxes represent write and read operations, respectively. Arrows denote the read-from relation. For simplicity, some operations are not shown in these figures. 
}
\label{f:abd}
\end{figure}

In this example, we assume that Darren is also Alice's friend, and can read all posts on Alice's wall.
Figure \ref{f:abd} shows an undesired result where Darren observes the posts in the order of ``{\bf lost}'', ``{\bf glad}'', and ``{\bf found}''. As addressed in Section \ref{s:intro}, this scenario is possible if the system only enforces eventual consistency. On the contrary, a system supporting intra-causal consistency would \textit{prevent} this scenario from happening. Let $H$ be the history shown in Figure \ref{f:abd}. Observe that intra-process dependency graph $D_D$, which shows that $r_D(Alice\_wall){\bf found} \precede{}{intra-CC}  r_D(x_{Alice\_wall}){\bf glad}$ -- this is due to the intra-program order specified by $D_D$. However, the timeline of Darren (in Figure \ref{f:abd}) shows that Darren observes ``{\bf glad}'' before observing ``{\bf found}''. This violates the intra-causal order, and hence, the scenario shown in Figure \ref{f:abd} should not occur in a system ensuring intra-causal consistency.


\end{itemize}

\subsubsection{Discussion}
\label{ss:disc_inter}

\paragraph{Construction of Intra-Process Dependency Graphs:}

In the discussion above, we have assumed that intra-process dependency graphs are given. However, in the real implementation, we need to specify how to construct the intra-process dependency graphs in an online fashion, i.e., when an operation $o$ is performed at a process $i$, which nodes should have edges to node $o$ in $D_i$? Generally, there are two approaches:

\begin{itemize}
\item Application at a process $i$ explicitly adds edges to node $o$ from some of the nodes corresponding to previous operations, since the application has knowledge of the semantic

\item We can also have a generic set of rules for adding edges for the given application. We will discuss some reasonable rules for Facebook-like application below. A complete set of rules is a future research direction. In the discussion below, let a {\em topic} depict a set of a single post and all the comments ensuing the post. Recall that posts and comments are defined in Section \ref{s:intro}. For example, in Figure \ref{f:alice}, ``{\bf lost}'' is a post, and ``{\bf No}'' is its comment. Moreover, ``{\bf lost}'' and ``{\bf No}'' form a topic, and ``{\bf found}'' and ``{\bf glad}'' form another topic. We also assume that given a history, we can infer the topic for each post or comment. In practice, this can be easily achieved by including a field representing the topic in the propagating message. Let us call operations corresponding to \textit{posts} in the local history $L_i$ as {\em post-operations}. Similarly, operations corresponding to {\em comments} is called {\em comment-operations}. Then, Facebook-like application should have some of the following rules:

\begin{itemize}
\item If operation $o$ is a comment-operation, then node $o$ should have an edge from some other node corresponding to an operation (either comment- or post-operation) that belongs to the same topic.

\item If operation $o$ is a post-operation, then node $o$ should have an edge from some other node corresponding to some post-operation, unless $o$ has no preceding operation.
\end{itemize}

The first rule effectively says that for a comment, users care about the order between the comment and other post or comments that belong to the same topic, whereas the second rule says that users care about the order among post-operations, but not necessarily the order from comment-operation to post-operation if they do not belong to the same topic.

Now, let us consider more complex application behavior: \textit{add or remove friends}. Suppose in the social network application, only friends can access the posts or comments on a user's wall. This is typically implemented via a separate checking mechanism that implements \textit{access control} at each replica. Suppose that user $j$ wants to read the most recent update from other user $i$, then checking mechanism at user $j$'s replica will check the friends graph (which is stored in each replica, as well), and propagate user $i$'s updates to user $j$ if users $i$ and $j$ are friends. We assume that friends graph is an undirected graph, i.e., friend relation is a mutual relation.

When the add/remove friend operations are interleaved with read and write operations, we need to carefully design the dependency graph to reflect the intended semantic. Consider a simple example: \textit{Alice first removes her boss Bob, and then posts that she wants a new job}. To ensure the expected application behavior, all later posts by Alice should not be observed by Bob after Alice removes Bob. This can be achieved by
treating the ``remove-Bob operation'' as a \textit{post-operation} addressed above. Due to the way that we constructed the intra-process dependency graph $D_A$, Alice's post looking for a job has a directed edge from ``remove-Bob operation''. Furthermore, by the \textit{transitivity rule}, \textit{intra-causal order} ensures that Bob would not be able to see any of Alice's later posts. This is because by the time when Bob's replica tries to make Alice's updates visible to Bob, the replica would have already received ``remove-Bob operation'' due to the enforced intra-causal order and updated the friends graph accordingly. Therefore, since Bob is not Alice's friend anymore, the checking mechanism will disallow Bob to read Alice's updates. Add friend operation can be treated similarly.

\end{itemize}

\paragraph{Potential Implementation:}

We briefly discuss how to implement a geo-replicated storage system that supports \textit{intra-causal consistency}. Our potential implementation is based on COPS \cite{Causal+,Causal+_ACMQ}, which implements a scalable causal memory for geo-replication. COPS has the following two key components:

\begin{itemize}
\item When an user issues a write operation $o_w$, the user-side library propagate the value along with a \textit{dependency tree} $T_w$ that keeps track of all operations that precedes $o_w$ defined by \textit{causal order} to the closest
available replica.

\item When an user issues a read operation, the closest available replica returns the value of the most recent write $o_w$ at the replica's local storage such that all of the operations in $T_w$ have already been observed by the user.
\end{itemize}
Please refer to \cite{Causal+,Causal+_ACMQ} for other details regarding COPS, e.g., how to maintain or propagate the dependency tree efficiently.

One modification we need for our system is to construct the \textit{dependency tree} according to \textit{intra-causal order}. More precisely, for a write operation $o_w$, the \textit{dependency tree} in our system will track all those operation that precedes $o_w$ defined by \textit{intra-causal order}. Please refer to \cite{thomas} for one implementation for a simple Facebook-like application.

\subsection{Other Intra-Consistency Models}
\label{ss:intra}

This section presents well-known consistency models and weakens them using intra-process dependency graphs. The weakened models are called intra-consistency models. 

\paragraph{Total Order:}

We start with consistency models that require the existence of a \textit{total order} among all processes. 

\begin{itemize}
\item {\em Sequential Consistency}:

\begin{definition}
\label{def:SC}
{\bf (Sequential Consistency\cite{lamport_sequential})}
A history $H$ is {\em sequentially consistent} if there is a serialization $S$ of $H$ that respects the program order $\precede{L_i}{}$ for each $p_i \in \sP$. 
\end{definition}

Recall that we have defined intra-program order $\precede{L_i}{intra}$ in Section \ref{s:intra}.

\begin{definition}
\label{def:intra-SC}
{\bf (Intra-Sequential Consistency)}
Given an intra-process dependency graph, a history $H$ is {\em intra-sequentially consistent} if there is a serialization $S$ of $H$ that respects the intra-program order $\precede{L_i}{intra}$ for each $p_i \in \sP$. 
\end{definition}

\item {\em Linearizability}: The second model is linearizability, which impose a real time constraint. A history $H$ induces a real time partial order $\precede{}{RT}$ as follows: whenever in $H$, the response of an operation $o_1$ {\em occurs before} the invocation of an operation $o_2$ in real time, then $o_1 \precede{}{RT} o_2$.

\begin{definition}
\label{def:L}
{\bf (Linearizability \cite{linearizability})}
A history $H$ is {\em linearizable} if there is a serialization $S$ of $H$ such that (i) $S$ respects the program order $\precede{L_i}{}$ for each $p_i \in \sP$; and (ii) $S$ respects the real time partial order $\precede{}{RT}$.
\end{definition}

Similarly, we can define the intra-real-time order $\precede{}{intra-RT}$ as follows: 

\begin{itemize}
\item For $o_1$ and $o_2$ that occur at two different processes in $H$, if the response of an operation $o_1$ {\em occurs before} the invocation of an operation $o_2$ in real time, $o_1 \precede{}{intra-RT} o_2$.
\end{itemize}

\begin{definition}
\label{def:intra-L}
{\bf (Intra-Linearizability)}
Given an intra-process dependency graph, a history $H$ is {\em intra-linearizable} if there is a serialization $S$ of $H$ such that (i) $S$ respects $\precede{L_i}{intra}$ for each $p_i \in \sP$; and (ii) $S$ respects the intra-real-time order $\precede{}{intra-RT}$.
\end{definition}

Intuitively, intra-linearizability requires (i) operations within the same process respect the intra-program order, and (ii) operations across different processes respect the real time partial order, whereas traditional linearizability requires all operations to respect the real time partial order. We believe such weakening is useful for the scenario of multi-thread executions at a single process.

\end{itemize}

\paragraph{Partial Order:}

Now, we consider models that only require a {\em partial order}. Note that causal consistency belongs to this category as well. 

\begin{itemize}
\item {\em PRAM Consistency}:

\begin{definition}
\label{def:PRAM}
{\bf (PRAM Consistency \cite{PRAM})}
A history $H$ is {\em PRAM consistent} if for each $p_i \in \sP$, there exists a serialization $S_i$ of $H|(p_i+W)$ that respects the program order $\precede{L_i}{}$.
\end{definition}

\begin{definition}
\label{def:intra-PRAM}
{\bf (Intra-PRAM Consistency)}
Given an intra-process dependency graph, a history $H$ is {\em Intra-PRAM consistent} if for each $p_i \in \sP$, there exists a serialization $S_i$ of $H|(p_i+W)$ that respects $\precede{L_i}{intra}$.
\end{definition}
\end{itemize}

%% file: FGB.tex
\section{Inter-Process Dependency}
\label{s:inter}

Our second approach takes into account {\bf where} the events take place. The notion is related to the ideas proposed in prior work \cite{fisheye,vector-field_consistency,Dagstuhl} -- observe events ``closer'' to you in a more consistent way, and observe ``farther'' events in a less consistent way. Below, we use the notion of \textit{inter-process dependency graph} to relax causal consistency to yield a model that we call \textit{inter-causal consistency}. Note that such relaxation may be categorized as (CC, EC)-fisheye consistency (Causal Consistency and Eventual Consistency) in \cite{fisheye}, but such a model is not discussed in \cite{fisheye}. Also, as we will show later, there are multiple different but reasonable ways to define {\em inter-causal consistency}. Moreover, we allow inter-process dependency graphs to be directed graphs, whereas \cite{fisheye} only considers undirected graphs. Thus, it is not clear how Friedman et al. \cite{fisheye} would define (CC, EC)-fisheye consistency. 

\begin{definition}
\label{def:inter}
{\bf (Inter-Process Dependency Graph)} A graph $G(\sv, \se)$ is an inter-process dependency graph if each node in $\sv$ corresponds to an unique process in $\sP$.
\end{definition}

Intuitively, given an application-specific parameter $d$, an inter-process dependency graph $G(\sv, \se)$ implies that the ordering of events at node $i \in \sv$ is important to node $j \in \sv$ if there exists a directed path of at most $d$-hop from $i$ to $j$ in $G$. Formally, an inter-process dependency graph $G$, a parameter $d$, and a history $H$ together induce the \textit{inter-causal order} (denoted $\precede{}{inter-CC}$). For simplicity, we ignore the inter-dependency graphs and $d$ in the notation. $o_1 \precede{}{inter-CC} o_2$ if any of the following holds:

\begin{itemize}
\item {\bf Program-order}: $o_1 \precede{L_i}{} o_2$ for some $p_i$ ($o_1$ precedes $o_2$ in $L_i$),

\item {\bf Inter-reads-from}: $o_1 \precede{}{read} o_2$ {\bf and} there exists a directed path with length at most $d$ from $user(o_1)$ to $user(o_2)$ in the inter-process dependency graph $G$, where $user(o)$ denotes the user performing operation $o$ ($o_2$ returns the value written by $o_1$ and $user(o_1)$ is $d$ hops away from $user(o_2)$ in $G$), or

\item {\bf Transitivity}: there is some other operation $o'$ such that $o_1 \precede{}{intra-CC} o' \precede{}{intra-CC} o_2$.
\end{itemize}

\begin{definition}
\label{def:inter-CC}
{\bf (Inter-Causal Consistency)}
A history $H$ is {\em inter-causally consistent} if for each $p_i \in \sP$, there exists a serialization $S_i$ of $H|(p_i+W)$ that respects $\precede{H}{inter-CC}$.
\end{definition}

\subsection{Discussion}
\label{ss:disc_intra}

\paragraph{Other Inter-Process Graphs:}

Facebook-like application's semantic implies that if two users $i, j$ are friends, then user $i$ would want to get an update from user $j$, and vice versa. Thus, it is reasonable to use friend's graph -- an undirected graph -- as the inter-process dependency graph. However, for other kinds of social network applications like Twitter or Pinterest, the semantic is different. Twitter-like or Pinterest-like application adopts a subscription-based semantic: user $i$ \textit{follows} or {\em subscribes} user $j$ if user $i$ is interested in learning updates from $j$; however, user $j$ may not necessarily want to learn user $i$'s update if user $j$ does not subscribe user $i$. Thus, it is more intuitive to use a directed graph to model such a subscription-based semantic. For Twitter-like application, we may use the \textit{subscription graph} -- which describes the subscription relations -- as the intra-process dependency graph.

\paragraph{Other Definitions of Inter-Causal Order:}

We only discuss one way of using inter-process dependency graph above. We propose two other methods that may be useful in some applications:

\begin{itemize}
\item Different types of operations performed by the users may be assigned a different distance parameter $d$.
Thus, for some type of operations, a causal ordering in a ``larger universe'' may be enforced, as compared
to other operations.
For instance, we may assign a larger universe size (i.e., larger $d$) to operations
that change the friend relationships.

\item We can also exploit the multiplicity of the directed paths. Denote by $m$ the given multiplicity parameter. For example, consider the friends graph in Facebook-like application, and let $d = 1$. Then, it may be reasonable to define the new {\em reads-from rule} as follows:

~

\noindent (i) $o_1 \precede{}{read} o_2$, (ii) $user(o_1)$ and $user(o_2)$ are friends or $user(o_1)$ and $user(o_2)$ share at least $m$ common friends. 

~

This kind of rule is reasonable in Facebook-like application -- even if $user(o_1)$ is not a neighbor of $user(o_2)$, $user(o_2)$ is still interested in learning $user(o_1)$'s update because they share enough common friends.
\end{itemize}

\paragraph{Potential Implementation:}

Our potential implementation is based on the implementation of causal shared memory presented in \cite{causal_memory}, which relies on using vector timestamps \cite{lamport_causal} to decide the most recent values. The algorithm is presented in \cite{alec}. How to make the implementation scalable for geo-replicated storage system is a future research direction.

\section{Summary}

This paper presents two methods to systematically weaken well-known consistency models. In particular, we discuss how to use two different kinds of \textit{dependency graphs} to weaken causal consistency, and why such relaxed models yield performance benefit while still being useful for social network applications.

%% file: appendix.tex
\section{Limitations of Eventual and Causal Consistency}
\label{a:drawback}

This section discusses some limitations of eventual and causal consistency in the context of social network applications. We will refer to an example in Figure \ref{f:alice} in Section \ref{s:intro}. Recall that in the scenario, there are three users, Alice, Bob and Calvin. Figure \ref{f:alice} shows Alice's wall, which contains the posts (by Alice) and comments (by Bob) and the corresponding timestamps. In the discussion, we will use the term {\em post} to refer to the text appearing first for a certain \textit{topic}, and the term {\em comment} for the text ensuing some {\em posts}. For example, Alice's update on 9:00 is a post, and Bob's update on 9:05 is an ensuing comment. For the example in Figure \ref{f:alice}, we will use ``{\bf lost}'' to represent Alice's post at 9:00, ``{\bf no}'' for Bob's comment at 9:05, ``{\bf found}'' for Alice's post at 9:30, and ``{\bf glad}'' for Bob's comment at 9:40, respectively.

\begin{figure}[hbtp!]
\centering
\begin{minipage}[b]{0.48\linewidth}
\includegraphics[scale=0.31]{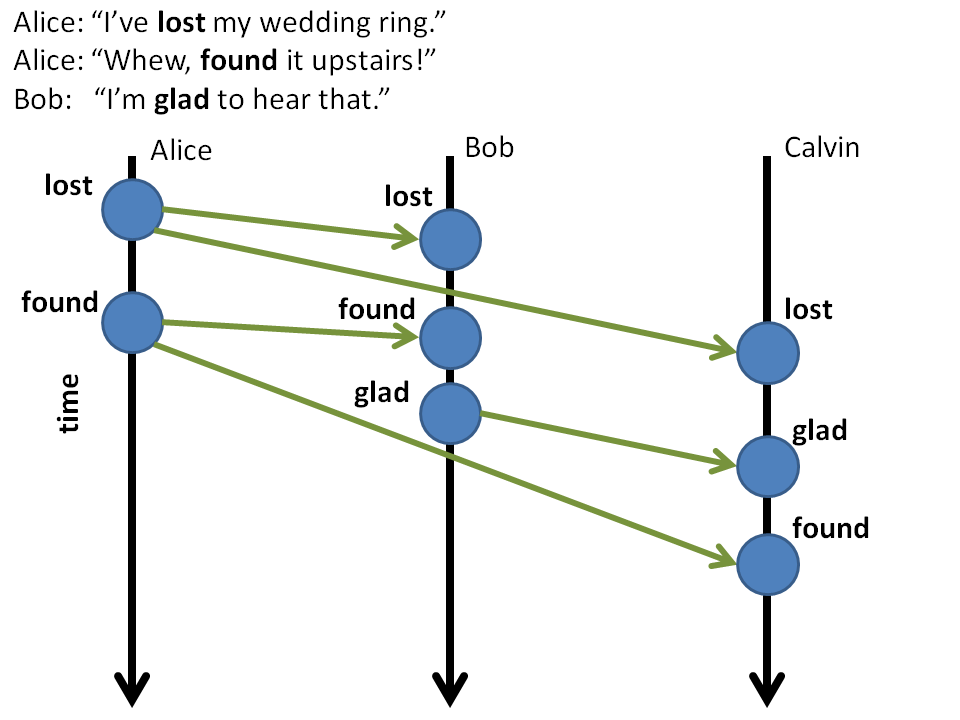}
\caption{\it Unfortunate outcome with eventual consistency \cite{Causal+_ACMQ}.
}
\label{fig:causal_bad}
\end{minipage}
\quad
\begin{minipage}[b]{0.48\linewidth}
\includegraphics[scale=0.31]{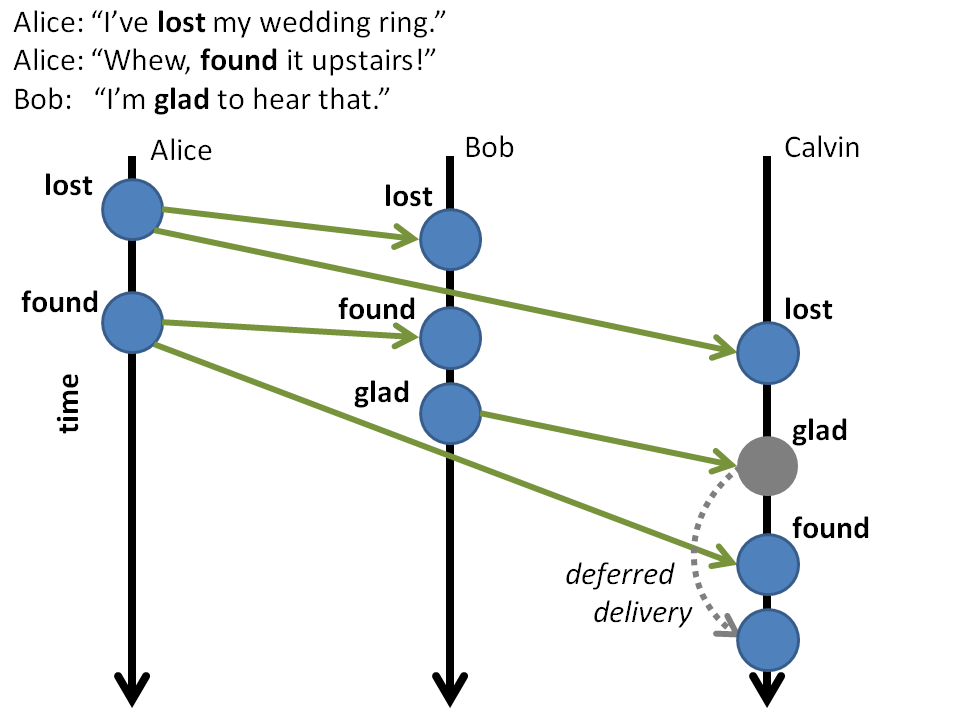}
\caption{\it Causal consistency achieves desired outcome
\cite{Causal+_ACMQ}.
}
\label{fig:causal}
\end{minipage}
\end{figure}

If the system only enforces {\em eventual consistency} \cite{Cassandra,Cassandra_paper}, then due to variable communication delays, Calvin may observe ``{\bf lost}'' and then ``{\bf glad}'' before having observed ``{\bf found}''. As a result, it will
appear to Calvin that Bob is glad to hear that Alice lost her wedding ring!
This
may occur if delay in propagating ``{\bf found}'' is large, as illustrated in
Figure \ref{fig:causal_bad}. In this figure, we ignore the propagation of ``{\bf no}'' for brevity.
Note that, as discussed above, the users interact with each other through the replicas of the geo-replicated storage.
For simplicity, these replicas are not shown in the figure, and we only show the outcome of the interactions.
For instance, the reason Calvin may observe ``{\bf found}'' before ``{\bf glad}'' may be that the delay in propagating
``{\bf found}'' to the replica accessed by Calvin is much larger than the delay in propagating ``{\bf glad}'' to that replica.  
Recall that the geo-replicated storage system is built upon an asynchronous communication network, so such a scenario may occur. 

If the system enforces {\em causal consistency} \cite{causal_memory}, then this unfortunate situation would never occur.
Instead, as depicted in Figure \ref{fig:causal}, Calvin will not observe ``{\bf glad}'' until he is able
to observe ``{\bf found}'' due to the enforced causal order \cite{lamport_causal}.
The implementation of causally consistent storage system \cite{Causal+,bolt-on} essentially achieves this desired outcome by
requiring the replica accessed by Calvin to \textit{delay} propagating ``{\bf glad}''
to Calvin until the replica receives the update corresponding to ``{\bf found}'' as shown in Figure \ref{fig:causal}.  Consequently, this may result into unnecessary latency for some events due to the restrictive ordering constraint of causal consistency model. The discussion of first method in Section \ref{s:intro} illustrate this drawback using an example.

